\documentclass{article}

\usepackage[final]{nips_2017}
\usepackage{graphicx}
\usepackage[utf8]{inputenc} 
\usepackage[T1]{fontenc}    
\usepackage{hyperref}       
\usepackage{url}            
\usepackage{booktabs}       
\usepackage{amsfonts}       
\usepackage{nicefrac}       
\usepackage{microtype}      

\usepackage{natbib}

\title{Parkinson's Disease Digital Biomarker Discovery with Optimized Transitions and Inferred Markov Emissions}

\author{
  Avinash Bukkittu \\
  Columbia University\\
  New York, NY 10027 \\
  \texttt{ab4377@columbia.edu} \\
 \And
  Baihan Lin \\
  Columbia University \\
  New York, NY 10027  \\
  \texttt{baihan.lin@columbia.edu} \\
   \And
  Trung Vu\\
  Columbia University \\
  New York, NY 10027  \\
  \texttt{ttv2107@columbia.edu} \\
   \AND
  Itsik Pe'er \\
  Columbia University \\
  New York, NY 10027  \\
  \texttt{itsik@cs.columbia.edu} \\
}

\begin{document}

\maketitle

\begin{abstract}

We search for digital biomarkers from Parkinson’s Disease by observing approximate repetitive patterns matching hypothesized step and stride periodic cycles. These observations were modeled as a cycle of hidden states with randomness allowing deviation from a canonical pattern of transitions and emissions, under the hypothesis that the averaged features of hidden states would serve to informatively characterize classes of patients/controls. We propose a Hidden Semi-Markov Model (HSMM), a latent-state model,  emitting 3D-acceleration vectors. Transitions and emissions are inferred from data. We fit separate models per unique device and training label. 

\textbf{Emission:} Our HSMM emits vectors which are normally distributed with per-state parameters of the respective Gaussian distributions. Unfortunately, these emitted vectors are not directly observed as the input data. Specifically, available data consists of relative coordinates, which change their reference coordinates as the device tilts across time points. We thus rotated the data to obtain absolute coordinates. These are consistent across timepoints for a sample, but might vary from sample to sample. We thus further detected the forward-backward direction for a given sample and oriented data along it and the unambiguous up-down axis. 

\textbf{Transition:} Hidden Markov Models (HMM) force geometric distributions of the duration spent at each state before transition to a new state. Instead, our HSMM allows us to specify the distribution of state duration. This modified version is more effective because we are interested more in the each state's duration than the sequence of distinct states, allowing inclusion of these durations the feature vector. 

\textbf{Featurization:} We extracted features from 1) means and standard deviations of time durations of each step, thus quantifying speed (based on velocity integration), pace, their irregularity; 2) the difference between the RecID frequency (f$\_$stride and f$\_$step) with consensus frequency (fc$\_$stride and fc$\_$step) in each of the 4 labels based on the frequency (Fourier) analysis, thus quantifying left-right asymmetry and walking speed; and 3) dwell times and state time fraction based on the HSMM, to capture the difference of walking style. 

\textbf{Augmentation:} We considered three options for data augmentation: 1) treating the outbound and inbound portions of the measurement together to double the data length for training; 2) flipping acceleration datasets along the left-right axis; and 3) trimming data points from the beginnings and ends of measurements. Missing data-points were replaced by imputed data based on the most probable label.

\end{abstract}
\section{Introductions}

\textbf{Motivation:} We observed approximate repetitive patterns matching hypothesized step and stride periodic cycles. We modeled this as a cycle of hidden states with randomness allowing deviation from a canonical pattern of transitions and emissions.  We hypothesize that the averaged features of hidden states would serve to informatively characterize classes of patients/controls. 

\textbf{Methodology: }We use an HSMM that emits 3D-acceleration vectors which are normally distributed with per-state parameters. Transitions and emissions are inferred from data per category. A category is defined by a unique device and label (before, after, on, or without medication).

\textbf{Novelty:}  Using relative coordinates can result in a lot of noise since the coordinates change from time point to time point. Using absolute coordinates provided by the accelerometer is better, but absolute coordinates might vary from sample to sample.  Our approach detects the forward-backward directions and converts the coordinate system of all samples to match their respective forward-backward directions. The HSMM framework allows us to specify the distribution of state duration instead of a traditional left-right HMM that only allows for geometric state durations.

\section{Methods}

\subsection{Preprocessing of the Accelerometer Datasets}

From the DREAM challenge \cite{dream}, the raw datasets of the accelerometer information given in a coordinate system which is relative to the phone, and therefore incomparable. Fortunately, the datasets of Device Motion actually offered us useful information of the orientation of the phone which can potentially be used to standardize the coordinate of the accelerometer datasets. We performed a rotation to change the reference frame of the coordinate system from the device to an absolute coordinate system. 

Next, we transform the absolute coordinate system to a coordinate system where x-axis always points in the forward walking direction of the subject, y-axis in the sideways direction, and z-axis downwards. To do this, we need to find the forward direction. From the device motion data, we have the gravity vector, which always points downward -- this is our z-axis. Therefore, the other two directions (forward and sideways) must lie in the plane perpendicular to the gravity vector. We assumed that the forward direction must be the direction in which the subject walks the farthest, and that this forward direction stays constant throughout the duration of the sample. For each sample, we find the direction that maximizes the distance walked (by integrating the projected acceleration in that direction twice) and select that direction as the forward direction. Thus, we now have the z-axis (gravity), the x-axis (direction that maximizes distance traveled), and y-axis (direction perpendicular to the other two). We then projected the acceleration vectors onto these 3 axes to get our desired coordinate system.

\subsection{Initial Hidden Markov Model based on the data}

\begin{figure}
  \includegraphics[scale=0.75]{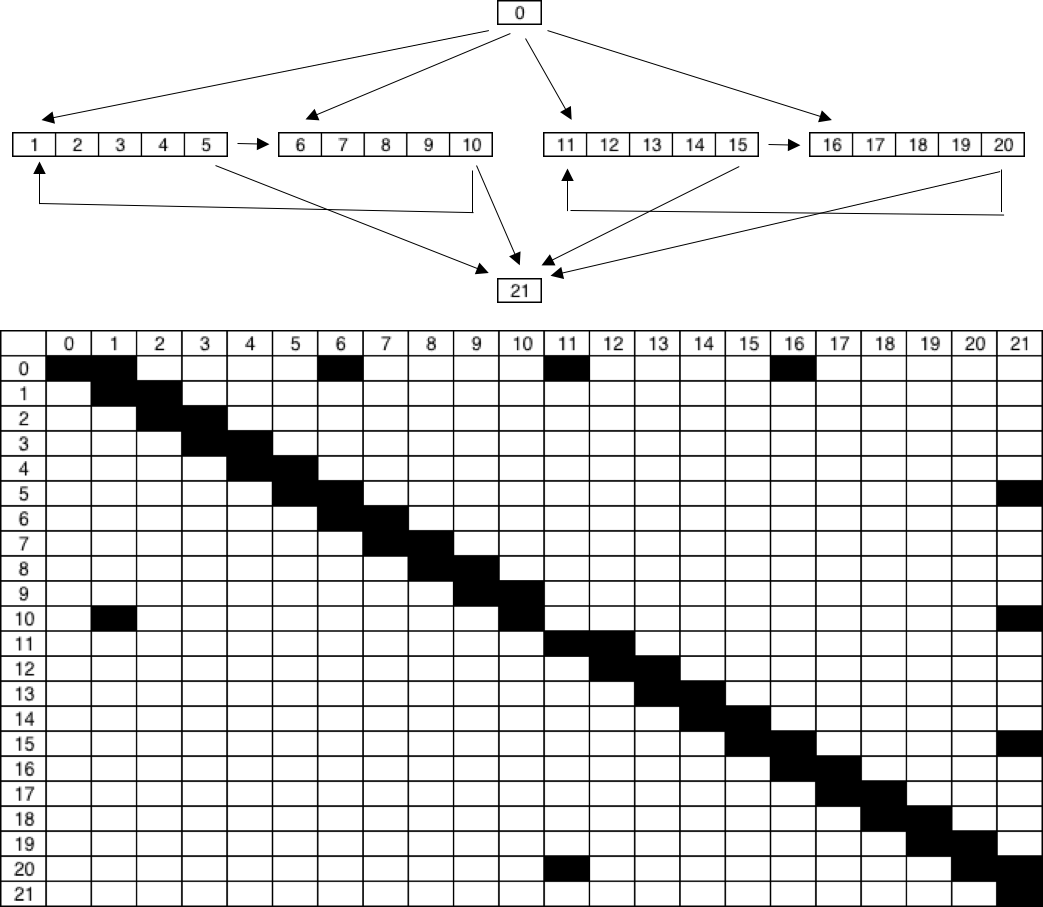}
  \caption{Diagram and transition matrix for our proposed Hidden Markov Model to capture the shape of walking with a premise that there are 10 states within each stride, about 5 states within each step, and the data were collected from subjects who put his/her cell phone in either left or right pocket and steps with either left or right foot first. State 0 is the initial resting stage and state 21 is the ending resting stage.}
  \label{fig:one}
\end{figure}

As shown in the transition matrix for Hidden Markov Model in Figure \ref{fig:one}, our proposed Hidden Markov Model captures the shape of walking with the premise that there are 10 states within each stride, about 5 states within each step. We also assume that the original raw data were collected from subjects who put his/her cell phone in either left or right pocket and step with either left or right foot first. Since state 0 is the initial resting stage and state 21 is the ending resting state, we assume there are four possible starting states for the periodic cycle: 1) phone in the right pocket and left foot  steps first; 2) phone in the right pocket and right foot  steps first; 3) phone in the left pocket and left foot  steps first; and 4) phone in the left pocket and right foot  steps first, corresponding to state 1, state 6, state 11 and state 16. We hypothesized that the Gaussian Mixture Hidden Markov Model, where the acceleration vectors go through a cycle of states, each state emitting acceleration vectors with a normal distribution, might be the proper implementation method.

\subsection{Hierarchical Dirichlet Process Hidden Semi-Markov Model}

\begin{figure}
  \includegraphics[scale=0.25]{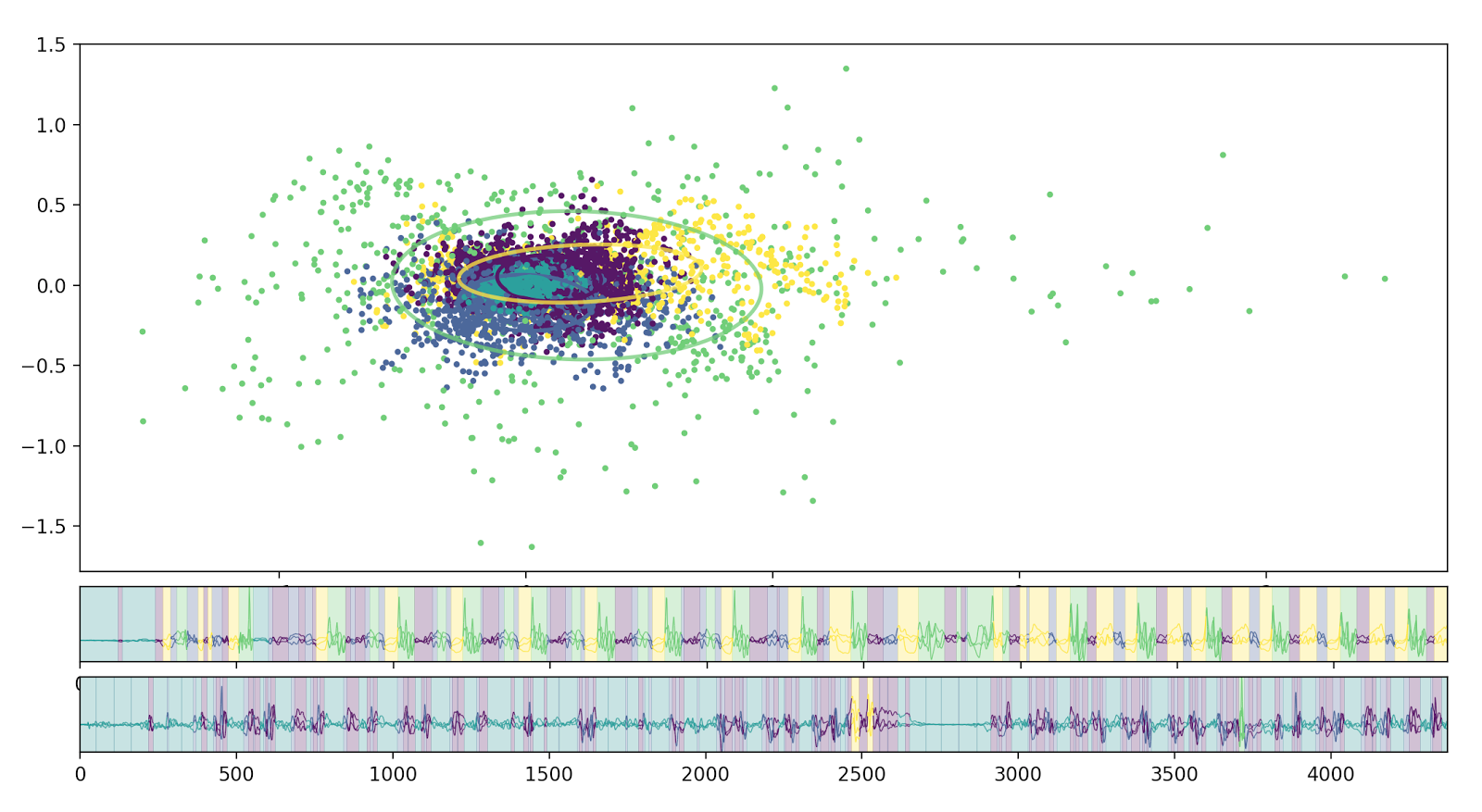}
  \caption{Visualization of the result from a test-run of the HPD-HSMM on two samples. The upper image shows the acceleration vectors clustered by states. The lower image shows where a state is entered through out the model.}
  \label{fig:two}
\end{figure}

A limitation of our initial model is that: (1) it assumes the duration that the subject stays in a given state is geometric, a memoryless distribution, (2) it assumes a fixed number of states, which might be inappropriate since it is only a hypothetical number and gait might vary from person to person. Thus, we decided to adapt an explicit duration Hidden semi-Markov modeling since we are interested more in the each state's duration given their explicit distribution instead of their sequences. We use these times for construction of our features. With the PyHSMM package by \cite{johnson2014bayesian}, we generated the sequences of unique hidden states as well as their dwell time as shown in Figure \ref{fig:two}. The generative process of this Hierarchical Dirichlet Process Hidden Semi-Markov Model (HDP-HSMM) is described as below in \cite{johnson2014bayesian}:

\includegraphics[scale=0.75]{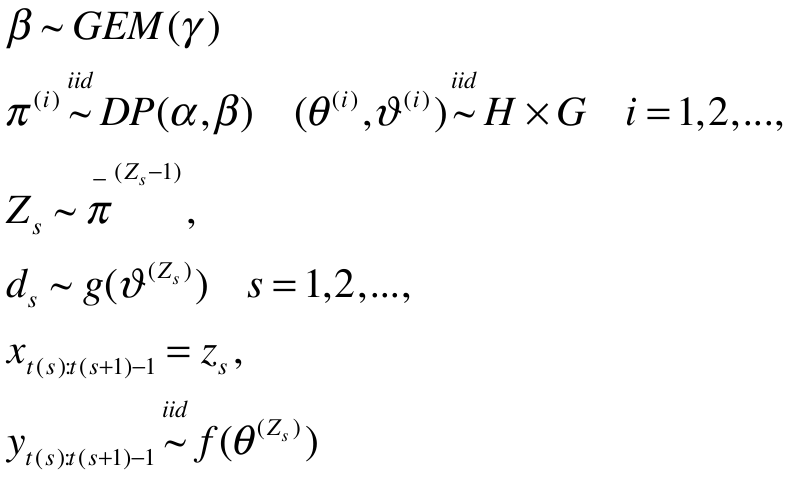}
  
\subsection{Data Augmentation}

Data augmentation is very useful to improve classification accuracy in different tasks. Similarly, we wish to explore how data augmentation can offer more information to our digital biomarker discovery. We proposed three data augmentation options: 1) treat the outbound and inbound together to double the data length for the training, 2) flip the acceleration datasets by the y-axis to generate a left-right symmetrical datasets, 3) trim data points in the very beginning or the ends. There are many considerations on how we should utilize these data augmentation options, for example, the inbound and outbound coordinate systems should be the same and both would contribute features towards the record ID. Therefore, the separation may be meaningful in terms of styles of gait potentially changing between these two activities. Our final set concatenated features from outbound data, from inbound data, and from the combined outbound and inbound data. For data trimming, we observed the ending timestamps should be absorbed in the ending states which in the best case may actually be informative. The flipping of the data helps the analysis by including plentiful of information to the minority group (probably the lefties), and that's meaningful in a competitive evaluation. Taking left as the default direction, we flipped every training and testing datasets before our model generations, to erase the noise caused by cancellation effect on y-axis by left-foot first versus right-foot first. 

\subsection{Feature Extraction}

For each recordId, we used the models that matched the recordID’s device category and predicted the Viterbi path. We extracted two important features from the Viterbi path. First, we collapsed the sequence such that adjacent hidden states with same values were merged into a single state. We then computed dwell time in the collapsed sequence of states. Second, we computed the fraction of time spent in each of the hidden states. We appended these two lists of features as the feature set. We performed this step for each matching models.

\subsection{Missing Data}

We noticed there are many cases, both in the training datasets and the testing datasets, which have missing outbound, or inbound, or both raw data. For these missing information, during the training stage, we skipped all these missing spaces. However, for the testing stage, we decided that replacing the blank space with the data generated from the most probable label will hurt the prediction the least. We divided the entire datasets into 16 groups: 4 labels (another time, before medication, after medication, and not taking medication) X 4 devices(iPhone6, iPhone5[GSM], iPhone5[GSM+CDMA],Others). We chose these devices based on their distribution in the datasets. We generated consensus outbound and inbound features for each groups which were used as features for missing data.

\section{Conclusions}

In summary, we extracted the following feature sets: 1) based on the velocity dataset generated from the preprocessing part, we calculated standard deviation and mean of time durations of each steps, aiming to find pace irregularity and walking speed; 2) based on the frequency analysis, we extracted the difference between the RecID frequency (f$\_$stride and f$\_$step) with consensus frequency (fc$\_$stride and fc$\_$step) in 4 labels, aiming to find left-right asymmetry and walking speed; 3) based on the HSMM, we generated the Viterbi path for each RecID by the models trained by each of the four labels, and extracted their dwell times in each stage, normalized into a fraction of sum 1, as well as their time duration spent in each unique hidden states of the Viterbi path. 

We also found that data cleaning is a very crucial step in this study. While the m-Power app collected plentiful information from a relatively large sample, the standardizations of the raw datasets are not guaranteed perfectly. Missing data, not fully followed instructions, noise from different devices and phone usage, as well as certain characteristics of walking irrelevant to the Parkinson’s Disease, all posed many challenges into this task. 

In the future, it would be very useful if we can further explore how different types of data augmentation methods can be further characterized and integrated into our algorithm. For example, currently the flipped information for left-right symmetry issue mentioned above, has not been fully accommodated due to the challenge of identifying which foot the subject step out first. If we can introduce this identification, we may increase our classification accuracy by training a more specialized model for each directions of walking.

\subsubsection*{Acknowledgements and Statements}

All authors characterized the materials, conceived the method, analyzed the data and wrote the manuscript. We acknowledge the valuable discussion and helpful advice from Pe’er Lab members, especially Ryan Bernstein, to our study. Source codes of this study can be accessed at \href{https://github.com/ab4377/dream-project}{https://github.com/ab4377/dream-project}.

\bibliographystyle{ACM-Reference-Format}
\bibliography{RSG_2017}


\begin{thebibliography}{2}


\ifx \showCODEN    \undefined \def \showCODEN     #1{\unskip}     \fi
\ifx \showDOI      \undefined \def \showDOI       #1{#1}\fi
\ifx \showISBNx    \undefined \def \showISBNx     #1{\unskip}     \fi
\ifx \showISBNxiii \undefined \def \showISBNxiii  #1{\unskip}     \fi
\ifx \showISSN     \undefined \def \showISSN      #1{\unskip}     \fi
\ifx \showLCCN     \undefined \def \showLCCN      #1{\unskip}     \fi
\ifx \shownote     \undefined \def \shownote      #1{#1}          \fi
\ifx \showarticletitle \undefined \def \showarticletitle #1{#1}   \fi
\ifx \showURL      \undefined \def \showURL       {\relax}        \fi
\providecommand\bibfield[2]{#2}
\providecommand\bibinfo[2]{#2}
\providecommand\natexlab[1]{#1}
\providecommand\showeprint[2][]{arXiv:#2}

\bibitem[\protect\citeauthoryear{Johnson et~al\mbox{.}}{Johnson
  et~al\mbox{.}}{2014}]%
        {johnson2014bayesian}
\bibfield{author}{\bibinfo{person}{Matthew~James Johnson} {et~al\mbox{.}}}
  \bibinfo{year}{2014}\natexlab{}.
\newblock {\em \bibinfo{title}{Bayesian time series models and scalable
  inference}}.
\newblock \bibinfo{thesistype}{Ph.D. Dissertation}.
  \bibinfo{school}{Massachusetts Institute of Technology}.
\newblock


\bibitem[\protect\citeauthoryear{syn8717496}{syn8717496}{2017}]%
        {dream}
\bibfield{author}{\bibinfo{person}{syn8717496}.}
  \bibinfo{year}{2017}\natexlab{}.
\newblock \bibinfo{title}{Parkinson's Disease Digital Biomarker DREAM
  Challenge}.
\newblock   (\bibinfo{year}{2017}).
\newblock


\end{thebibliography}

\end{document}